\documentclass{article} 

\usepackage{stmaryrd}
\usepackage{bussproofs}
\EnableBpAbbreviations 
\newcommand{\UICm}[1]{\UIC{$#1$}}
\newcommand{\AXCm}[1]{\AXC{$#1$}}
\newcommand{\BICm}[1]{\BIC{$#1$}}
\newcommand{\TICm}[1]{\TIC{$#1$}}
\usepackage{amsmath,amssymb}

\def\dtop{{\dot{\top}}}
\def\dand{{\dot{\wedge}}}
\def\tildeimp{\tilde{\Rightarrow}}

\def\tildeand{\tilde{\wedge}}
\def\tildeor{\tilde{\vee}}
\def\tildetop{\tilde{\top}}
\def\tildebot{\tilde{\bot}}
\def\tildefa{\tilde{\forall}}
\def\tildeex{\tilde{\exists}}
\def\impc{\check{\Rightarrow}}
\def\andc{\check{\wedge}}
\def\orc{\check{\vee}}
\def\topc{\check{\top}}
\def\botc{\check{\bot}}
\def\fac{\check{\forall}}
\def\exc{\check{\exists}}
\def\imp{\Rightarrow}
\def\and{\wedge}

\def\fa{\forall}
\def\ex{\exists}
\def\calB{{\mathcal{B}}}
\def\calM{{\mathcal{M}}}
\def\calD{{\mathcal{D}}}

\def\calS{{\mathcal{S}}}
\def\calT{{\mathcal{T}}}
\def\lra{\longrightarrow}

\def\calA{{\mathcal{A}}}

\def\calE{{\mathcal{E}}}

\def\calL{{\mathcal{L}}}

\newcommand{\interp}[1]{{\llbracket #1 \rrbracket}}

\newcommand{\couic}[1]{}

\newbox\tempa
\newbox\tempb
\newdimen\tempc
\def\mud#1{\hfil $\displaystyle{\mathstrut #1}$\hfil}
\def\rig#1{\hfil $\displaystyle{#1}$}
\def\irulehelp#1#2#3{\setbox\tempa=\hbox{$\displaystyle{\mathstrut #2}$}%
                        \setbox\tempb=\vbox{\halign{##\cr
        \mud{#1}\cr
        \noalign{\vskip\the\lineskip}
        \noalign{\hrule height 0pt}
        \rig{\vbox to 0pt{\vss\hbox to 0pt{${\; #3}$\hss}\vss}}\cr
        \noalign{\hrule}
        \noalign{\vskip\the\lineskip}
        \mud{\copy\tempa}\cr}}
                      \tempc=\wd\tempb
                      \advance\tempc by \wd\tempa
                      \divide\tempc by 2 }
\def\irule#1#2#3{{\irulehelp{#1}{#2}{#3}
                     \hbox to \wd\tempa{\hss \box\tempb \hss}}}
\def\birulehelp#1#2#3{\setbox\tempa=\hbox{$\displaystyle{\mathstrut #2}$}%
                        \setbox\tempb=\vbox{\halign{##\cr
        \mud{#1}\cr
        \noalign{\vskip\the\lineskip}
        \noalign{\hrule height 0pt}
        \phantom{$#3$}
                \rig{\vbox to 0pt{\vss\hbox to 0pt{${\; #3}$\hss}\vss}}\cr
        \noalign{\hrule}
        \noalign{\vskip\the\lineskip}
        \mud{\copy\tempa}\cr}}
                      \tempc=\wd\tempb
                      \advance\tempc by \wd\tempa
                      \divide\tempc by 2 }
\def\birule#1#2#3{{\birulehelp{#1}{#2}{#3}
                     \hbox to \wd\tempa{\hss \box\tempb \hss}}\phantom{#3}}

\newtheorem{definition}{Definition}{}
\newtheorem{proposition}{Proposition}{}
\newtheorem{theorem}{Theorem}{}
{}
\newenvironment{proof}{\noindent {\em Proof.}}{}
\newenvironment{remark}{\noindent {\em Remark.}}{}

\title{A Simple Proof That Super-Consistency\\
Implies Cut Elimination}
\author{Gilles Dowek\thanks{INRIA,
    23 avenue d'Italie, CS 81321, 75214 Paris Cedex 13, France,
{\tt gilles.dowek@polytechnique.edu},
{\tt http://www-roc.inria.fr/who/Gilles.Dowek/}.}
~and Olivier Hermant
\thanks{ISEP,
21 rue d'Assas, 75006 Paris, France,
{\tt olivier.hermant@isep.fr},
{\tt http://perso.isep.fr/ohermant}.
}}

\date{}

\begin{document}
\maketitle

\begin{abstract}
We give a simple and direct proof that super-consistency implies the
cut elimination property in deduction modulo. This proof can be seen
as a simplification of the proof that super-consistency implies proof
normalization. It also takes ideas from the semantic proofs of cut
elimination that proceed by proving the completeness of the cut-free
calculus. As an application, we compare our work with the cut
elimination theorems in higher-order logic that involve V-complexes.
\end{abstract}

\section{Introduction}
{\em Deduction modulo} is an extension of predicate logic where some
axioms may be replaced by rewrite rules.  For instance, the axiom $x +
0 = x$ may be replaced by the rewrite rule $x + 0 \lra x$ and the
axiom $x \subseteq y \Leftrightarrow \fa z~(z \in x \Rightarrow z \in
y)$ by the rewrite rule $x \subseteq y \lra \fa z~(z
\in x \Rightarrow z \in y)$.

In the model theory of Deduction modulo, it is important to
distinguish the fact that some propositions 
are computationally equivalent, {\em i.e.} congruent
({\em e.g.} $x \subseteq y$ and $\fa z~(z
\in x \Rightarrow z \in y)$), in which case they should have the same
value in a model, from the fact that they are provably equivalent, 
in which case
they may have different values. This has lead, in \cite{GDow06}, to the
introduction of a generalization of Heyting algebras called {\em truth values
algebras} and a notion of $\calB$-valued model, where $\calB$ is a
truth values algebra.  We have called {\em super-consistent} the
theories that have a $\calB$-valued model for all truth values
algebras $\calB$ and we have given examples of super-consistent
theories as well as examples of consistent theories that are not
super-consistent.

In Deduction modulo, there are theories for which there exist proofs
that do not normalize.  But, we have proved in \cite{GDow06} that all
proofs normalize in super-consistent theories.  This proof
proceeds by observing that reducibility candidates \cite{Girard} can
be structured in a truth values algebra and thus that super-consistent
theories have reducibility candidate valued models. Then, the
existence of such a model implies proof normalization \cite{GDowBWer03} and
hence cut elimination. As many theories, in particular arithmetic and
simple type theory, are super-consistent, we get Gentzen's and
Girard's theorems as corollaries.

This paper is an attempt to simplify this proof replacing the algebra
of reducibility candidates ${\mathcal{C}}$ by a simpler truth values
algebra ${\mathcal{S}}$. Reducibility candidates are sets of proofs. We
show that we can replace each proof of such a set by its conclusion,
obtaining this way sets of sequents, rather than sets of proofs, for
truth values. 

Although the truth values of our model are sets of sequents, our cut
elimination proof uses another truth values algebra whose elements are
sets of contexts: the algebra of contexts $\Omega$, that happens to be
a Heyting algebra. From any ${\mathcal{S}}$-valued model of a theory we
build a second-level model, that is $\Omega$-valued and that we use to
show cut-elimination.

This technique gives a proof that uses ideas taken from both methods
employed to prove cut elimination: proof-term normalization and
completeness of the cut-free calculus. From the first, come the ideas
of truth values algebra and neutral proofs and from the second, the
idea of building a model such that sequents valid in this model have
cut-free proofs.

This paper is an extended version of a conference paper
\cite{GDowOHer07} by the same authors. Some technical inaccuracies of
\cite{GDowOHer07} have been corrected in this version. In Section
\ref{sec:sc} we recall the technical material that will be 
useful to understand Section \ref{sec:ce}, that is the core of the
paper. At the end of the paper, we provide an analysis of the proof
obtained in the case of higher-order logic and compare it with other
semantic proofs.

\section{Super-Consistency} \label{sec:sc}
To keep the paper self contained, we recall in this section the
definition of Deduction modulo, truth values algebras, $\calB$-valued
models and super-consistency. A more detailed presentation can be
found in \cite{GDow06}.

\subsection{Deduction Modulo}

Deduction modulo \cite{DHK,GDowBWer03} is an extension of predicate logic
(either single-sorted or many-sorted predicate logic) where a theory
is defined by a set of axioms $\Gamma$ and a congruence $\equiv$,
itself defined by a confluent rewrite system rewriting terms to terms
and atomic propositions to propositions.

In this paper we consider natural deduction rules. These rules are
modified to take the congruence $\equiv$ into account.  For example,
the elimination rule of the implication is not formulated
as usual
$$\irule{\Gamma \vdash A \Rightarrow B~~~\Gamma \vdash A}{\Gamma
\vdash B}{}$$
but as
$$\irule{\Gamma \vdash C~~~\Gamma \vdash A}{\Gamma \vdash B}{C \equiv
A \Rightarrow B}$$ 
All the deduction rules are modified in a similar way as shown in
Figure \ref{fig:dn-mod}. Note that the usual proviso that $x$ does
not appear freely in the $\forall_i$ and $\exists_e$ rules holds, as
informally reminded by the side condition. See \cite{GDowBWer03} for a
more thorough presentation.

\begin{figure}[t]
\leftskip -52pt
\begin{tabular}{|cc|}
\hline
~&~\\
\multicolumn{2}{|c|}{
\AXCm{}
\RightLabel{axiom ($B \equiv A$ for some $A$ in $\Gamma$)}
\UICm{\Gamma \vdash B}
\DP
}\\
~&~\\
\AXCm{\Gamma, A \vdash B}
\RightLabel{$\Rightarrow_i$, $C \equiv A \Rightarrow B$}
\UICm{\Gamma \vdash C}
\DP
& 
\AXCm{\Gamma \vdash C}
\AXCm{\Gamma \vdash A}
\RightLabel{$\Rightarrow_e$, $C \equiv A \Rightarrow B$}
\BICm{\Gamma \vdash B}
\DP
\\
~&~\\
\AXCm{\Gamma \vdash A}
\AXCm{\Gamma \vdash B}
\RightLabel{$\wedge_i$, $C \equiv A \wedge B$}
\BICm{\Gamma \vdash C}
\DP
& 
\AXCm{\Gamma \vdash C}
\RightLabel{$\wedge_{e}$-1, $C \equiv A \wedge B$}
\UICm{\Gamma \vdash A}
\DP
\\
~&~ \\
&
\AXCm{\Gamma \vdash C}
\RightLabel{$\wedge_e$-2, $C \equiv A \wedge B$}
\UICm{\Gamma \vdash B}
\DP
\\
~&~\\
\AXCm{\Gamma \vdash A}
\RightLabel{$\vee_i$-1, $C \equiv A \vee B$}
\UICm{\Gamma \vdash C}
\DP
& \\
~&~ \\
\AXCm{\Gamma \vdash B}
\RightLabel{$\vee_i$-2, $C \equiv A \vee B$}
\UICm{\Gamma \vdash C}
\DP
&
\AXCm{\Gamma, A \vdash C}
\AXCm{\Gamma, B \vdash C}
\AXCm{\Gamma \vdash D}
\RightLabel{$\vee_e$, $D \equiv A \vee B$~~}
\TICm{\Gamma \vdash C}
\DP
\\
~&~\\
\AXCm{\phantom{\Gamma \vdash B}}
\RightLabel{$\top_i$, $A \equiv \top$}
\UICm{\Gamma \vdash A}
\DP
& 
\AXCm{\Gamma \vdash B}
\RightLabel{$\bot_e$, $B \equiv \bot$}
\UICm{\Gamma \vdash A}
\DP
\\
~&~\\
\AXCm{\Gamma \vdash A}
\RightLabel{$\forall_i$, $B \equiv \forall{x} A$, with $x \notin FV(\Gamma)$}
\UICm{\Gamma \vdash B}
\DP
&
\AXCm{\Gamma \vdash B}
\RightLabel{$\forall_e$, $B \equiv \forall{x} A$ and $C \equiv (t/x)A$}
\UICm{\Gamma \vdash C}
\DP
\\
~&~\\
\AXCm{\Gamma \vdash C}
\RightLabel{$\ex_i$, $B \equiv \exists{x} A$ and $C \equiv (t/x) A$}
\UICm{\Gamma \vdash B}
\DP
& 
\AXCm{\Gamma \vdash C}
\AXCm{\Gamma, A \vdash B}
\RightLabel{$\exists_e$, $C \equiv \exists{x} A$, with $x \notin
  FV(\Gamma, B)$}
\BICm{\Gamma \vdash B}
\DP \\
~&~\\
\hline
\end{tabular}
\caption{Rules of Natural Deduction Modulo}
\label{fig:dn-mod}
\end{figure}

In Deduction modulo, there are theories for which there exist proofs
that do not normalize.
For instance, in the theory formed with the rewrite rule 
$P \lra (P \Rightarrow Q)$, the proposition $Q$ has a proof 
$$\hspace*{-1.25cm}
\irule{\irule{\irule{\irule{}
                             {P \vdash P \Rightarrow Q}
                             {\mbox{axiom~~}}
                      \;\;\;\;\;\;\;\;\;\;\;\;\;\;
                      \irule{}
                            {P \vdash P}
                            {\mbox{axiom~~}}
                      }
                      {P \vdash Q}
                      {\mbox{$\Rightarrow$-elim~~}}
               }
               {\vdash P \Rightarrow Q}
               {\mbox{$\Rightarrow$-intro}}
         \;\;\;\;\;\;\;\;\;\;\;\;\;\;\;\;\;\;\;\;\;\;\;\;\;\;\;\;\;\;\;\;\;\;\;\;\;\;\;\;\;\;\;\;\;\;
         \irule{\irule{\irule{}
                             {P \vdash P \Rightarrow Q}
                             {\mbox{axiom~~}}
                      \;\;\;\;\;\;\;\;\;\;\;\;\;\;
                      \irule{}
                            {P \vdash P}
                            {\mbox{axiom~~}}
                      }
                      {P \vdash Q}
                      {\mbox{$\Rightarrow$-elim~~}}
               }
               {\vdash P}
               {\mbox{$\Rightarrow$-intro}}
       }
       {\vdash Q}
       {\mbox{$\Rightarrow$-elim}}$$
that does not normalize.
In some other theories, such as the theory
formed with the rewrite rule $P \lra (Q \Rightarrow P)$, all proofs
strongly normalize.

In Deduction modulo, like in predicate logic, closed normal proofs
always end with an introduction rule.  Thus, if a theory can be
expressed in Deduction modulo with rewrite rules only, {\em i.e.} with
no axioms, in such a way that proofs modulo these rewrite rules
strongly normalize, then the theory is consistent, it has the
disjunction property and the witness property, and various proof search
methods for this theory are complete.

Many theories can be expressed in Deduction modulo with rewrite rules
only, in particular arithmetic and simple type theory, and the notion
of cut of Deduction modulo subsumes the notions of cut defined for
each of these theories.  For instance, simple type theory can be
defined as follows.

\begin{definition}[Simple type theory \cite{Church40,DHK-HOL,LIP03}]\label{def:stt}
The sorts are inductively defined:
\begin{itemize}
\item $\iota$ and $o$ are sorts,
\item if $T$ and $U$ are sorts then $T \rightarrow U$
is a sort.
\end{itemize}
The language contains the constants $K_{T,U}$ of sort $T \rightarrow U
\rightarrow T$,
$S_{T,U,V}$ of sort $(T \rightarrow U \rightarrow V) \rightarrow (T
\rightarrow U) \rightarrow T \rightarrow V$,
$\dot{\top}$ of sort $o$ and $\dot{\bot}$ of sort $o$,
$\dot{\Rightarrow}$, $\dot{\wedge}$ and $\dot{\vee}$ 
of sort $o \rightarrow o \rightarrow o$, 
$\dot{\fa}_{T}$ and $\dot{\ex}_T$ of sort $(T \rightarrow o) \rightarrow o$,
the function symbols
$\alpha_{T,U}$ of rank $\langle T \rightarrow U, T, U \rangle$ 
and the predicate symbol
$\varepsilon$ of rank $\langle o \rangle$.

The rules are
$$\alpha(\alpha(\alpha(S_{T,U,V},x),y),z)  \longrightarrow  \alpha(\alpha(x,z),\alpha(y,z))$$
$$\alpha(\alpha(K_{T,U},x),y)  \longrightarrow  x$$
$$\varepsilon(\dot{\top}) \longrightarrow \top$$
$$\varepsilon(\dot{\bot}) \longrightarrow \bot$$
$$\varepsilon(\alpha(\alpha(\dot{\Rightarrow},x),y)) \longrightarrow 
\varepsilon(x) \Rightarrow \varepsilon(y)$$ 
$$\varepsilon(\alpha(\alpha(\dot{\wedge},x),y)) \longrightarrow \varepsilon(x) \wedge \varepsilon(y)$$ 
$$\varepsilon(\alpha(\alpha(\dot{\vee},x),y)) \longrightarrow \varepsilon(x) \vee \varepsilon(y)$$ 
$$\varepsilon(\alpha(\dot{\fa}_T,x)) \longrightarrow \fa y~\varepsilon(\alpha(x,y))$$
$$\varepsilon(\alpha(\dot{\ex}_T,x)) \longrightarrow \ex y~\varepsilon(\alpha(x,y))$$

\end{definition}

\subsection{Truth Values Algebras}\label{sec:tva}

\begin{definition}[Truth values algebra]
\label{tva}
Let $B$ be a set, whose elements are called 
{\em truth values}, $B^{+}$ be a subset of $B$, 
whose elements are called {\em positive} truth values,
$\calA$ and $\calE$ be subsets of $\wp(B)$,
$\tildetop$ and $\tildebot$ be elements of $B$, $\tildeimp$,
$\tildeand$, and $\tildeor$ be functions from $B \times B$ to
$B$, $\tildefa$ be a function from $\calA$ to $B$ and
$\tildeex$ be a function from $\calE$ to $B$.  
The structure 
$\calB = \langle B, B^{+}, \calA, \calE, \tildetop, \tildebot,
\tildeimp, \tildeand, \tildeor, \tildefa, \tildeex\rangle$ is said to
be {\em a truth values algebra} if the set 
$B^{+}$ is {\em closed under the intuitionistic deduction rules} 
{\em i.e.} if for all $a$, $b$, $c$ in $B$, $A$ in $\calA$
and $E$ in $\calE$,
\begin{enumerate}
\item if $a~\tildeimp~b \in B^{+}$ and $a \in B^{+}$ then 
$b \in B^{+}$, 
\item $a~\tildeimp~b~\tildeimp~a \in B^{+}$, 
\item
$(a~\tildeimp~b~\tildeimp~c)~\tildeimp~(a~\tildeimp~b)~\tildeimp~a~\tildeimp~c
\in B^{+}$, 
\item
$\tildetop \in B^{+}$, 
\item $\tildebot~\tildeimp~a \in B^{+}$, 
\item $a~\tildeimp~b~\tildeimp~(a~\tildeand~b) \in B^{+}$, 
\item $(a~\tildeand~b)~\tildeimp~a \in B^{+}$,
\item $(a~\tildeand~b)~\tildeimp~b \in B^{+}$,
\item $a~\tildeimp~(a~\tildeor~b) \in B^{+}$,
\item $b~\tildeimp~(a~\tildeor~b) \in B^{+}$,
\item
$(a~\tildeor~b)~\tildeimp~(a~\tildeimp~c)~\tildeimp~(b~\tildeimp~c)~\tildeimp~  c \in B^{+}$, 
\item the set $a~\tildeimp~A = \{a~\tildeimp~e~|~e \in A\}$ is in
$\calA$ and\\ the set $E~\tildeimp~a = \{e~\tildeimp~a~|~e \in E\}$ is in
$\calA$, 
\item if all elements of $A$ are in $B^{+}$ 
then $\tildefa~A \in B^{+}$, 
\item 
$\tildefa~(a~\tildeimp~A)~\tildeimp~a~\tildeimp~(\tildefa~A) \in 
B^{+}$, 
\item if $a \in A$, then $(\tildefa~A)~\tildeimp~a \in B^{+}$, 
\item if $a \in E$, then $a~\tildeimp~(\tildeex~E) \in B^{+}$, 
\item 
$(\tildeex~E)~\tildeimp~\tildefa~(E~\tildeimp~a)~\tildeimp~a \in 
B^{+}$.
\end{enumerate}
\end{definition}

\begin{proposition}
\label{Heyting}
Any Heyting algebra is a truth values algebra.  The operations
$\tildetop$, $\tildeand$, $\tildefa$ are greatest lower bounds, the
operations $\tildebot$, $\tildeor$, $\tildeex$ are least upper bounds,
the operation $\tildeimp$ is the arrow of the Heyting algebra, and
$B^{+} = \{\tildetop\}$.
\end{proposition}

\proof{See \cite{GDow06}}. 

\begin{definition}[Fullness]
A truth values algebra is said to be {\em full} if $\calA = \calE =
\wp(B)$, {\em i.e.} if $\tildefa~A$ and $\tildeex~A$ exist for all
subsets $A$ of $B$.
\end{definition}

\begin{definition}[Ordered truth values algebra]
An {\em ordered truth values algebra} is a 
truth values algebra together with a relation 
$\sqsubseteq$ on $B$ such that 
\begin{itemize}
\item $\sqsubseteq$ is an order relation, {\em i.e.} a reflexive,
  antisymmetric and transitive relation,
\item $B^{+}$ is upward closed,
\item $\tildetop$ and $\tildebot$ are maximal and minimal elements,
\item $\tildeand$, $\tildeor$, 
$\tildefa$ and $\tildeex$ are monotone, 
$\tildeimp$ is left anti-monotone and right monotone.
\end{itemize}
\end{definition}

\begin{definition}[Complete ordered truth values algebra]
An ordered truth values algebra is said to be {\em complete} if
every subset of $B$ has a greatest lower bound for $\sqsubseteq$.
\end{definition}

\subsection{Models}

\begin{definition}[$\calB$-structure] \label{def:Bstructure}
Let $\calL = \langle f_i, P_j \rangle$ be a language for predicate
logic and $\calB$ be a truth values algebra. A {\em $\calB$-structure}
$\calM = \langle M, \calB, \hat{f}_i,
\hat{P}_j \rangle$, for the language $\calL$, 
is a structure such that $\hat{f_i}$ is a function
from $M^n$ to $M$ where $n$ is the arity of the symbol $f_i$
and $\hat{P_j}$ is a function from $M^n$ to $B$ where $n$ is
the arity of the symbol $P_j$.

This definition extends trivially to many-sorted languages.
\end{definition}

\begin{definition}[Denotation] \label{def:denotation}
Let $\calB$ be a truth values algebra, $\calM$ be a $\calB$-structure
and $\phi$ be an assignment. The denotation of propositions and terms
in $\calM$ is defined inductively as follows:

\begin{itemize}
\item $\llbracket x \rrbracket_{\phi} = \phi(x)$, 
\item $\llbracket f(t_1, ..., t_n) \rrbracket_{\phi} = 
\hat{f}(\llbracket t_1\rrbracket_{\phi}, ..., \llbracket
t_n\rrbracket_{\phi})$,
\item $\llbracket P(t_1, ..., t_n) \rrbracket_{\phi} = 
\hat{P}(\llbracket t_1\rrbracket_{\phi}, ..., \llbracket
t_n\rrbracket_{\phi})$,
\item $\llbracket \top \rrbracket_{\phi} = \tildetop$, 
\item $\llbracket \bot \rrbracket_{\phi} = \tildebot$, 
\item $\llbracket A \Rightarrow B \rrbracket_{\phi} = 
\llbracket A \rrbracket_{\phi} 
~\tildeimp~
\llbracket B \rrbracket_{\phi}$, 
\item
$\llbracket A \wedge B \rrbracket_{\phi} = 
\llbracket A \rrbracket_{\phi}
~\tildeand~
\llbracket B \rrbracket_{\phi}$, 
\item
$\llbracket A \vee B \rrbracket_{\phi} = 
\llbracket A \rrbracket_{\phi}
~\tildeor~
\llbracket B \rrbracket_{\phi}$, 
\item $\llbracket \fa x~A \rrbracket_{\phi} = 
\tildefa~\{
\llbracket A \rrbracket_{\phi + (d/x)}~|~d \in M\}$,
\item $\llbracket \ex x~A \rrbracket_{\phi} = 
\tildeex~\{
\llbracket A \rrbracket_{\phi + (d/x)}~|~d \in M\}$.
\end{itemize}
Notice that the denotation of a proposition containing quantifiers may be
undefined, but it is always defined if the truth values algebra is
full.
\end{definition}

\begin{definition}[Denotation of a context and of a sequent]
\label{denotecontext}
The denotation $\llbracket A_1, ..., A_n \rrbracket_{\phi}$
of a context $A_1, ..., A_n$ is that of the proposition
$A_1 \wedge ... \wedge A_n$.
The denotation $\llbracket A_1, ..., A_n \vdash B \rrbracket_{\phi}$
of the sequent $A_1, ..., A_n \vdash B$ is that of the proposition
$(A_1 \wedge ... \wedge A_n) \Rightarrow B$.
\end{definition}

\begin{definition}[Model]\label{def:model}
A proposition $A$ is said to be {\em valid} in a $\calB$-structure
$\calM$, and the $\calB$-structure $\calM$ is said to be a {\em model
of} $A$ if for all assignments $\phi$,
$\llbracket A \rrbracket_\phi$ is defined and is a positive truth
value.

Consider a theory in Deduction modulo defined by a set of axioms $\Gamma$
and a congruence $\equiv$.
The
$\calB$-structure $\calM$ is said to be a model of the
theory $\Gamma, \equiv$ if all axioms of $\Gamma$ are valid in $\calM$
and for all terms or propositions $A$ and $B$ such that $A \equiv B$
and assignments $\phi$, $\llbracket A
\rrbracket_\phi$ and $\llbracket B \rrbracket_\phi$ are defined and
$\llbracket A \rrbracket_\phi = \llbracket B \rrbracket_\phi$.
\end{definition}

Deduction modulo is sound and complete with respect to this notion of
model.

\begin{proposition}[Soundness and completeness]
\label{soundcomplete}
The proposition $A$ is provable in the theory formed with the axioms 
$\Gamma$ and the congruence $\equiv$
if and only if 
it is valid in all the models of $\Gamma,\equiv$ where the truth values
algebra is full, ordered and complete.
\end{proposition}

\proof{See \cite{GDow06}.}

\subsection{Super-Consistency}

\begin{definition}[Super-consistent]
A theory in Deduction modulo formed with the axioms $\Gamma$ and the
congruence $\equiv$ is {\em super-consistent} if it has a
$\calB$-valued model for all full, ordered and complete truth values
algebras $\calB$.
\end{definition}

\begin{proposition}
\label{stt}
Simple type theory is super-consistent.
\end{proposition}

\proof{Let $\calB$ be a full truth values algebra.  We build the model
$\calM$ as follows.  The domain $M_{\iota}$ is any non empty set, for
instance the singleton $\{0\}$, the domain $M_{o}$ is ${\calB}$
and the domain $M_{T \rightarrow U}$ is the set
$M_U^{M_T}$ of functions from $M_T$ to $M_U$.
The interpretation of the symbols of the language is
$\hat{S}_{T,U,V} = a \mapsto (b \mapsto (c \mapsto a(c)(b(c))))$,
$\hat{K}_{T,U} = a \mapsto (b \mapsto a)$,
$\hat{\alpha}(a,b) = a(b)$,
$\hat{\varepsilon}(a) = a$,
$\hat{\dot{\top}} = \tildetop$, 
$\hat{\dot{\bot}} =\tildebot$,
$\hat{\dot{\Rightarrow}} = \tildeimp$, 
$\hat{\dot{\wedge}} = \tildeand$,
$\hat{\dot{\vee}} = \tildeor$,
$\hat{\dot{\fa}}_T = a \mapsto \tildefa(Range(a))$,
$\hat{\dot{\ex}}_T = a \mapsto \tildeex(Range(a))$
where $Range(a)$ is the range of the function $a$. The model $\calM$
is a $\calB$-valued model of simple type theory.}

\section{Cut Elimination} \label{sec:ce}
\label{sec:first}
\subsection{The Algebra Of Sequents}

\begin{definition}[Neutral proof]
A proof is said to be {\em neutral} if its last rule is the axiom rule
or an elimination rule, but not an introduction rule.
\end{definition}

We now define the notion of cut-free proof. Instead of giving a
syntactic definition (absence of cut) we give a positive inductive
definition.

\begin{definition}[Cut-free proofs]\label{def:cfproofs} 
{\em Cut-free} proofs are defined inductively as follows:
\begin{itemize}
\item a proof that ends with the axiom rule is cut-free, 
\item a proof that ends with an introduction rule and where the
premises of the last rule are proved with cut-free proofs is cut-free,
\item a proof that ends with an elimination rule and where the major
premise of the last rule is proved with a neutral cut-free proof and
the other premises with cut-free proofs is cut-free.
\end{itemize}
\end{definition}

\begin{definition}[The algebra of sequents] \label{def:candidates}~
\begin{itemize}
\item $\tildetop$ is the set of sequents $\Gamma \vdash C$ that have a
neutral cut-free proof or such that $C \equiv \top$.

\item $\tildebot$ is the set of sequents $\Gamma \vdash C$ that have a
neutral cut-free proof.

\item $a~\tildeand~b$ is the set of sequents $\Gamma \vdash C$ that
have a neutral cut-free proof or such that $C \equiv (A \wedge B)$
with $(\Gamma \vdash A) \in a$ and $(\Gamma \vdash B) \in b$.

\item $a~\tildeor~b$ is the set of sequents $\Gamma \vdash C$ that
have a neutral cut-free proof or such that $C \equiv (A \vee B)$ with
$(\Gamma \vdash A) \in a$ or $(\Gamma \vdash B) \in b$.

\item $a~\tildeimp~b$ is the set of sequents $\Gamma \vdash C$ that
have a neutral cut-free proof or such that $C \equiv (A \Rightarrow
B)$ and for all contexts $\Sigma$ such that $(\Gamma, \Sigma \vdash A)
\in a$, we have $(\Gamma, \Sigma \vdash B) \in b$.

\item $\tildefa~S$ is the set of sequents $\Gamma \vdash C$ that have
a neutral cut-free proof or such that $C \equiv (\fa x~A)$ and for every
term $t$ and every $a$ in $S$, $(\Gamma \vdash (t/x)A) \in a$.

\item $\tildeex~S$ is the set of sequents $\Gamma \vdash C$ that have
a neutral cut-free proof or such that $C \equiv (\ex x~A)$ and for
some term $t$ and some $a$ in $S$, $(\Gamma \vdash (t/x)A) \in a$.
\end{itemize}

Let $S$ be the smallest set of sets of sequents closed under
$\tildetop$, $\tildebot$, $\tildeand$, $\tildeor$, $\tildeimp$,
$\tildefa$, $\tildeex$ and by arbitrary intersections.
\end{definition}

\begin{proposition}
The structure $\calS = \langle S, S, \wp(S), \wp(S),
\tildetop, \tildebot, \tildeimp, \tildeand, \tildeor, \tildefa,
\tildeex, \subseteq \rangle$ is a full, ordered and complete truth
values algebra.
\end{proposition}

\proof{As all truth values are positive, the conditions of Definition 
\ref{tva} are obviously met. Thus $\calS$ is a truth values algebra. 
As the domains of $\tildefa$ and $\tildeex$ are defined as $\wp(S)$,
this algebra is full. As it is closed under arbitrary intersections, all
subsets of $S$ have a greatest lower bound, thus all subsets of $S$ have a least upper
bound and the algebra is complete.}

\medskip 
\noindent
{\em Remark.} The algebra $\calS$ is not a Heyting algebra. In
particular $\tildetop~\tildeand~\tildetop$ and 
$\tildetop$ are different: the first set contains the sequent 
$\vdash \top \wedge \top$, but not the second. 

\medskip

\begin{proposition} \label{prop:basics}
For all elements $a$ of $S$, contexts $\Gamma$, and propositions $A$
and $B$
\begin{itemize}
\item $(\Gamma, A \vdash A) \in a$,

\item if $(\Gamma \vdash B) \in a$ then $(\Gamma, A \vdash B) \in a$,

\item if $(\Gamma, A, A \vdash B) \in a$ then $(\Gamma, A \vdash B)
  \in a$,

\item if $(\Gamma \vdash A) \in a$ and $B \equiv A$ then $(\Gamma
\vdash B) \in a$,

\item if $(\Gamma \vdash A) \in a$ then $\Gamma \vdash A$ has a cut
free proof.

\end{itemize}
\end{proposition}

\proof{The first proposition is proved by noticing that the sequent
$\Gamma, A \vdash A$ has a neutral cut-free proof. The others are
  proved by a simple induction on the construction of $a$. For
  instance, if $a = c~\tildeand~d$, then:
\begin{itemize}
\item if $\Gamma \vdash B$ has a neutral cut-free proof, so has
  $\Gamma, A \vdash B$. Otherwise $B \equiv (C \wedge D)$,
  $(\Gamma \vdash C) \in c$ and $(\Gamma \vdash D) \in d$. By
  induction hypothesis,  $(\Gamma, A \vdash C) \in c$ and $(\Gamma, A
  \vdash D) \in d$ so by definition $\Gamma, A \vdash B \in c
 ~\tildeand~d$.
\item if $\Gamma, A, A \vdash B$ has a neutral cut-free proof, so has
  $\Gamma, A \vdash B$. Otherwise $B \equiv (C \wedge D)$,
  $(\Gamma, A, A \vdash C) \in c$ and $(\Gamma, A, A \vdash D) \in
  d$. By induction hypothesis,  $(\Gamma, A \vdash C) \in c$ and
  $(\Gamma, A \vdash D) \in d$ so by definition $\Gamma, A \vdash B
  \in c~\tildeand~d$.
\item if $\Gamma \vdash A$ has a neutral cut-free proof, so has
  $\Gamma \vdash B$. Otherwise $A \equiv (C \wedge D) \equiv B$,
  $(\Gamma \vdash C) \in c$ and $(\Gamma \vdash D) \in d$ so by
  definition $\Gamma \vdash B \in c~\tildeand~d$.
\item if $\Gamma \vdash A$ has a neutral cut-free proof there is
  nothing to show. Otherwise $A \equiv (C \wedge D)$,
  $(\Gamma \vdash C) \in c$ and $(\Gamma \vdash D) \in d$. By
  induction hypothesis $\Gamma \vdash C$ and $\Gamma \vdash D$ have
  cut-free proofs and we can add a $\wedge$-intro rule to obtain a
  cut-free proof of $\Gamma \vdash A$.
\end{itemize}
}

\medskip 

Consider a super-consistent theory $\Gamma, \equiv$. By definition, it
has a $\calS$-model $\calM$.  In the rest of the paper, $\calM$ refers
to this model. Its domain is written $M$.

\begin{proposition}[Substitution] \label{prop:substitution}
Let $A$ be a proposition, $\phi$ an assignment, $x$ a variable and $t,
u$ terms. Let $\phi' = \phi + (\interp{t}_\phi / x)$. Then
$\interp{(t/x)u}_\phi = \interp{u}_{\phi'}$ and $\interp{(t/x)A}_\phi
= \interp{A}_{\phi'}$.
\end{proposition}

\proof{By structural induction.}

\subsection{The Algebra Of Contexts}

\begin{definition}[Fiber]\label{fiber}
Let $b$ be a set of sequents, $A$ be a proposition, $\sigma$ be a
substitution and $f$ be a function mapping propositions to sets of
sequents. We define the parametrized fiber over $A$ in $b$, $b
\lhd_\sigma^{f} A$, as the set of contexts $\Gamma = A_1, \cdots, A_n$
such that for any $\Delta$ such that $(\Delta \vdash \sigma A_i) \in
f(A_i)$ for any $i$, we have $(\Delta \vdash \sigma A) \in b$.
\end{definition}

\begin{definition}[$\Gamma$-adapted context]\label{def:adapted}
Let $\Gamma = A_1, ..., A_n$ be a set of propositions, $\phi$ be an
assignment $\sigma$ be a substitution. Let $\Delta$ be a set of
propositions. We say that $\Delta$ is $\Gamma$-adapted for $\phi, \sigma$
(in short: $\Delta$ is $\Gamma$-adapted) if, for any $i$, $(\Delta
\vdash \sigma A_i) \in \interp{A_i}_\phi$. 
\end{definition}

\begin{proposition}[Composition of adapted contexts]\label{lem:adapt-composition}
Let $\Gamma_1, \Gamma_2$ be two sets of propositions. Let $\phi$ be an
assignment, $\sigma$ be a substitution and $\Delta_1, \Delta_2$ be
$\Gamma_1$-adapted (resp. $\Gamma_2$-adapted) contexts for
$\phi,\sigma$. Then,
\begin{itemize}
\item $\Delta_1, \Delta_2$ is $\Gamma_1, \Gamma_2$-adapted for $\phi,
\sigma$.
\item if $\Delta_1 = \Delta, B, B$ then $\Delta, B$ is
  $\Gamma_1$-adapted for $\phi, \sigma$.
\end{itemize}
\end{proposition}
\proof{Let $A$ be a member of $\Gamma_1$ and $A'$ be a member of
  $\Gamma_2$. $(\Delta_1 \vdash \sigma A) \in \interp{A}_\phi$ and
  $(\Delta_2 \vdash \sigma A') \in \interp{A'}_\phi$ by definition. Then,
\begin{itemize}
\item $(\Delta_1, \Delta_2 \vdash \sigma A) \in \interp{A}_\phi$ and
  $(\Delta_1, \Delta_2 \vdash \sigma A') \in \interp{A'}_\phi$ by the
  second point of Proposition \ref{prop:basics}.
\item if $\Delta_1$ matches the hypothesis, then $(\Delta, B \vdash
  \sigma A) \in \interp{A}_\phi$ by the third point of Proposition
  \ref{prop:basics}.
\end{itemize}
}

\begin{definition}[Outer value]\label{mod:outer}
Let $A$ be a proposition, we define the set of contexts $[A]$ as the 
set of contexts $\Gamma = A_1, \cdots, A_n$ such that for any
assignment $\phi$, any substitution $\sigma$, and any context
$\Delta$, whenever $(\Delta \vdash \sigma A_i) \in
\llbracket A_i \rrbracket_\phi$ for any $i$ (in other words, $\Delta$
is $\Gamma$-adapted), then $(\Delta \vdash \sigma A) \in \llbracket A
\rrbracket_\phi$. Note that $[A]$ is precisely $\llbracket A
\rrbracket \lhd^{\llbracket\_\rrbracket}_\sigma A$.
\end{definition}

\begin{proposition} \label{prop:basics2}
For any context $\Gamma$, any propositions $A$ and $B$,
\begin{itemize}
\item $(\Gamma, A) \in [A]$,

\item if $\Gamma \in [B]$ then 
$(\Gamma,A) \in [B]$,

\item if $(\Gamma, A, A) \in [B]$ then $(\Gamma, A) \in [B]$,

\item if $\Gamma \in [A]$ and $B \equiv A$ then 
$\Gamma \in [B]$,

\item if $\Gamma \in [A]$ then $\Gamma \vdash A$ has a cut-free proof.
\end{itemize}
\end{proposition}

\proof{ This follows directly from the definitions and
  Proposition \ref{prop:basics}. For the first point, consider some
  $\phi, \sigma$ and a $\Delta$ that is $(\Gamma, A)$-adapted for
  $\phi, \sigma$. In particular, by Definition \ref{mod:outer},
  $(\Delta \vdash \sigma A) \in 
  \interp{A}_\phi$, which was to be proved. The second point
  restricts the sets of contexts: if $\Delta$ is $(\Gamma,
  A)$-adapted then it is obviously $\Gamma$-adapted and the conditions
  of Definition \ref{mod:outer} are fulfilled. Similarly, for the
  third point, if $\Delta$ is $(\Gamma, A, A)$-adapted then it is
  $(\Gamma, A)$-adapted. The fourth point follows from
  Proposition \ref{prop:basics}, since $\interp{A}_\phi =
  \interp{B}_\phi$ by 
  definition of the model $\calM$. The last point is a consequence of
  Proposition \ref{prop:basics}.}

\begin{definition}[The algebra of contexts]
\label{def:ha}
Let $\Omega$ be the smallest set of sets of contexts containing all the 
$[A]$ for some proposition $A$ and closed under arbitrary intersections.
\end{definition}

\begin{remark}
Notice that an element $c$ of $\Omega$ can always be written in the form 
$$
c = \bigcap_{i \in \Lambda_c} [A_i]
$$
\end{remark}

\begin{proposition}
The set $\Omega$ ordered by inclusion is a complete Heyting algebra.
\end{proposition}

\proof{As $\Omega$ is ordered by inclusion and closed under
arbitrary intersections, the greatest lower bound of any subset of
$\Omega$ can be defined as the intersection of all its elements and it
always exists. Thus, all its subsets also have least upper bounds,
namely the greatest lower bound of its majorizers.

The operations $\topc$, $\andc$ and $\fac$ are
defined as nullary, binary and infinitary greatest lower bounds and
the operations $\botc$, $\orc$ and $\exc$ are
defined as nullary, binary and infinitary least upper bounds.
Finally, the arrow $\impc$ of two elements $a$ and $b$ is the least
upper bound of all the $c$ in $\Omega$ such that $a \cap c \leq b$
$$a~\impc~ b = \exc~ \{ c \in \Omega~|~a \cap c \leq b\}$$

To prove that $\Omega$ is a Heyting algebra, $\impc$ must have some
specific properties \cite{ATroDvDal88}. The following condition is
necessary and sufficient:
$$
a \leq b~\impc~c \mbox{ iff } a \cap b \leq c
$$
The reverse implication holds by elementary lattice theory from the
very definition of $\impc$, but the direct one requires some
work. Let $\Gamma = A_1, \cdots, A_n$ and assume it belongs to $a
\cap b$, our aim is to show that $\Gamma \in c = \bigcap [C_i]$. Let
$C$ be one of the $C_i$, we show that $\Gamma \in [C]$.

Unfolding the assumption, we know that $a \leq \exc {\sf D}$, with
${\sf D} = \{ d ~|~ b \cap d \leq c \}$, or, unfolding a step further
(see (\ref{eqn:two})):
\begin{eqnarray}
a & \leq & \bigcap \left\{ [E] ~|~ \bigcup{\sf D} \subseteq [E]
\right\} \label{eqn:one}
\end{eqnarray}
We first show that $\bigcup{\sf D} \subseteq [\Gamma \imp C]$,
where $\Gamma \imp C$ denotes $A_1 \imp \cdots A_n \imp C$, and later
take advantage of this in (\ref{eqn:one}).

Let $d \in {\sf D}$, and $\Delta \in d$. We have $(\Delta, \Gamma) \in
d$ and $(\Delta, \Gamma) \in a \cap b \subseteq b$ by Proposition
\ref{prop:basics2}. So, $(\Delta, \Gamma) \in b \cap d$, and since $d
\in {\sf D}$, $(\Delta, \Gamma) \in c \subseteq [C]$.

We now prove by induction on $n$ (the cardinality of $\Gamma$) that if
$(\Delta, \Gamma) \in [C]$ then $\Delta \in [\Gamma \imp C]$. This is
immediate if $n = 0$. Otherwise, we apply induction hypothesis, and we
get that $(\Delta, A_1) \in [\Gamma_1 \imp C]$. Let $\phi$ be an
assignment, $\sigma$ be a substitution and $\Sigma_\Delta$ be a
$\Delta$-adapted context. We must show that $(\Sigma_\Delta \vdash
\sigma \Gamma \imp \sigma C) = (\Sigma_\Delta \vdash \sigma A_1 \imp \sigma
\Gamma_1 \imp \sigma C)$ belongs to $\interp{\Gamma \imp C}_{\phi} =
\interp{A_1}_\phi \tildeimp \interp{\Gamma_1 \imp C}_\phi$. 

Let $\Sigma_{A_1}$ such that $\Sigma_{A_1} \vdash \sigma A_1 \in
\interp{A_1}_\phi$. $\Sigma_{A_1}$ is $A_1$-adapted and from
Proposition \ref{lem:adapt-composition} 
$\Sigma_\Delta, \Sigma_{A_1}$ is $\Delta, A_1$-adapted. So,
$\Sigma_\Delta, \Sigma_{A_1} \vdash \sigma \Gamma_1 \imp \sigma A \in
\llbracket \Gamma_1 \imp C \rrbracket_\phi$. This exactly means that
$\Sigma_\Delta \in \interp{A_1}_\phi \tildeimp \interp{\Gamma_1 \imp
  C}_\phi$, from Definition \ref{def:candidates} of $\tildeimp$.
Therefore, by the very Definition \ref{mod:outer}, $\Delta \in [A_1
  \imp \Gamma_1 \imp C]$.\\

So $[\Gamma \imp C]$ is an upper bound of ${\sf D}$ and, by
(\ref{eqn:one}), of $a$. By hypothesis on $\Gamma$, $\Gamma \in a
\cap b \subseteq a \subseteq [\Gamma \imp C]$.\\

Let us call $\Gamma_i$ the context $A_{i+1}, ..., A_n$. $\Gamma_i$ is
a suffix of $\Gamma$ and we show, by induction on $i \leq n$, that
$\Gamma \in [\Gamma_i \imp C]$. The base case has just been proved
above.

For the inductive step, we assume that $\Gamma \in [\Gamma_i \imp
  C] = [A_{i+1} \imp \Gamma_{i+1} \imp C]$. Let $\phi$ be an
assignment, $\sigma$ be a substitution and 
$\Delta_\Gamma$ be a $\Gamma$-adapted context. We show that
$(\Delta_\Gamma \vdash \sigma \Gamma_{i+1} \imp \sigma C) \in
\interp{\Gamma_{i+1} \imp C}_\phi$. This will allow to conclude that
$\Gamma \in [\Gamma_{i+1} \imp C]$.

$(\Delta_{\Gamma} \vdash \sigma A_{i+1} \imp \sigma \Gamma_{i+1} \imp
\sigma C) \in \llbracket A_{i+1} \imp \Gamma_{i+1} \imp C
\rrbracket_\phi$ by the induction hypothesis, so
$$
(\Delta_{\Gamma} \vdash \sigma A_{i+1} \imp \sigma \Gamma_{i+1} \imp
\sigma C) \in \interp{A_{i+1}}_\phi \tildeimp \interp{\Gamma_{i+1}
  \imp C}_\phi
$$

If the sequent $\Delta_\Gamma \vdash \sigma A_{i+1} \imp \sigma
\Gamma_{i+1} \imp \sigma C$ has a neutral cut-free proof, we add an
elimination rule with a cut-free proof of $\Delta_\Gamma \vdash \sigma
A_{i+1}$ (obtained by Proposition \ref{prop:basics} since
$(\Delta_\Gamma \vdash \sigma A_{i+1}) \in \interp{A_{i+1}}_\phi$). This
gives a neutral cut-free proof of the sequent $\Delta_{\Gamma} \vdash
\sigma \Gamma_{i+1} \imp \sigma C$ and this sequent therefore belongs
by Definition \ref{def:candidates} to $\interp{\Gamma_{i+1} \imp
  C}_\phi$.

Otherwise, following Definition \ref{def:candidates} of $\tildeimp$,
since $\Delta_{\Gamma} \vdash \sigma A_{i+1} \in
\interp{A_{i+1}}_\phi$ , we conclude directly that $\Delta_{\Gamma}
\vdash \sigma \Gamma_{i+1} \imp \sigma C \in \interp{\Gamma_{i+1} \imp
  C}_\phi$.\\

Consider now the $n^{\mbox{th}}$ case of the previous statement. It
states $\Gamma \in [C]$, which was to be proved. This holds for any of
the $C_i$ such that $c = \bigcap_{i \in \Lambda_c} 
[C_i]$, so $\Gamma$ belongs to their intersection $c$, and finally $a
\cap b \subseteq c$ is proved.}\\

The binary least upper bound, $a~\orc~b$, of $a$ and
$b$ is the intersection of all the elements of $\Omega$ that contain
$a \cup b$. From Definition \ref{def:ha}
$$
a~\orc~b = \bigcap_{(a \cup b) \subseteq c} c  =
\bigcap_{(a \cup b) \subseteq \bigcap
     [A_i]} (\bigcap
       [A_i]) = \bigcap_{(a \cup b) \subseteq
       [A]} [A]
$$
The infinitary least upper bound $\exc~E$ of the elements
of a set $E$ is the intersection of all the elements of $\Omega$ that
contain the union of the elements of $E$. For the same reason as
above
\begin{eqnarray}
& \exc~E = \bigcap_{(\bigcup E)~\subseteq~c} c = \bigcap_{(\bigcup E
  )~\subseteq~[A]} [A] & \label{eqn:two}
\end{eqnarray}

Notice that the nullary least upper bound $\botc$ is $\bigcap \{ a ~|~
d \leq a, \mbox{ for any } d \in \emptyset \}$ {\em i.e.} the
intersection of all the elements of $\Omega$. Also, the nullary
greatest lower bound $\topc$ {\em is} the set of all contexts, we show
in Proposition \ref{prop:chech} below that it is equal to $[\top] \in
\Omega$ and hence, that this construction is well-defined.

Finally, notice that $\Omega$ might be a non trivial Heyting algebra,
although the quotient Heyting algebra $\calS/S^+$ is always trivial
because $S^{+} = S$. The construction of Definition \ref{def:ha} does
then not boil down to this quotient and produces a more informative
structure.\\

The next proposition, the Key lemma of our proof, shows that the outer
values of compound propositions can be obtained from the outer values
of their components using the corresponding operation of the Heyting
algebra $\Omega$.  Notice that, unlike most semantic cut
elimination proofs \cite{MOka02,LIP03,OHerJLip08b}, we directly prove
equalities in this lemma, and not just inclusions, although the cut
elimination proof is not completed yet.
\begin{proposition}[Key lemma]
\label{prop:chech}
For all propositions $A$ and $B$
\begin{itemize}
\item $[\top] = \topc$,
\item $[\bot] = \botc$,
\item $[A \wedge B] = [A]~\andc~ [B]$,
\item $[A \vee B] = [A]~\orc~[B]$, 
\item $[A \Rightarrow B] =
[A]~\impc~[B]$,
\item $[\forall{x} A] = 
\fac~\{[(t/x)A]~|~ t \in \calT\}$, 
\item $[\exists{x} A] = 
\exc~\{ [(t/x)A]~|~ t \in \calT \}$.
\end{itemize}
where $\calT$ is the set of open terms in the language of the theory.
\end{proposition}

\proof{
\begin{itemize}
\item Let $\Gamma$ be a context, and $\Delta$ be a
  $\Gamma$-adapted context. By Definition \ref{def:candidates},
  $(\Delta \vdash \top) \in \tildetop = \interp{\top}$. Thus
  $\Gamma \in [\top]$, and $[\top] = \topc$.

\item The set $\botc$ is the intersection of all $[C]$. In 
particular, $\botc \subseteq [\bot]$. Conversely,
let $\Gamma \in [\bot]$, let $\phi$ be an assignment, $\sigma$ be a
substitution and $\Delta$ be $\Gamma$-adapted. Consider an
arbitrary $C$. By Definition \ref{def:candidates}, $\Delta
\vdash \bot$ has a neutral cut-free proof. So does $\Delta
\vdash \sigma  C$ and this sequent belongs to $\interp{C}_\phi$, thus
$\Gamma \in [C]$. Hence $\Gamma$ is an element of all $[C]$ and
therefore of their intersection $\botc$.

\item Let $\Gamma \in [A]~\andc~[B] = [A] \cap [B]$.  Let $\phi$ be an
  assignment, $\sigma$ be a substitution and $\Delta$ be
  $\Gamma$-adapted. We have $\Gamma \in [A]$ and $\Gamma \in [B]$ and
  thus $(\Delta \vdash \sigma A) \in \interp{A}_{\phi}$ and
  $(\Delta \vdash \sigma B) \in \interp{B}_{\phi}$. From
  Definition \ref{def:candidates}, we get $(\Delta \vdash \sigma (A
  \wedge B)) \in \interp{A \wedge B}_{\phi}$. Hence $\Gamma \in [A
    \wedge B]$.

  Conversely, let $\Gamma \in [A \wedge B]$, let $\phi$ be an
  assignment, $\sigma$ be a substitution and consider a
  $\Gamma$-adapted context $\Delta$. We have $(\Delta \vdash
  \sigma (A \wedge B)) \in
  (\interp{A}_{\phi}~\tildeand~\interp{B}_{\phi})$. If $\Delta \vdash
  \sigma (A \wedge B)$ has a neutral and cut free proof, then so do
  $\Delta \vdash \sigma A$ and $\Delta \vdash \sigma B$ by the
  $\wedge$-elim rules and this shows that $(\Delta \vdash \sigma A)
  \in \interp{A}_\phi$ and $(\Delta \vdash \sigma B) \in
  \interp{B}_\phi$. Otherwise, those last two statements follow
  directly from Definition \ref{def:candidates}. We conclude that
  $\Gamma \in [A]$ and $\Gamma \in [B]$, hence  that $\Gamma \in [A]
  \cap [B] = \Gamma \in [A]~\andc~[B]$.

\item To show $[A] ~\orc~ [B] \subseteq [A \vee B]$ it is sufficient
  to prove that $[A \vee B]$ is an upper bound of $[A]$ and
$[B]$. Let $\Gamma \in [A]$, let $\phi$ be an assignment, $\sigma$ be
  a substitution and $\Delta$ be $\Gamma$-adapted. By
  hypothesis, $(\Delta \vdash \sigma A) \in 
\interp{A}_{\phi}$ and by Definition \ref{def:candidates} this
means that $(\Delta \vdash \sigma (A \vee B)) \in
(\interp{A}_{\phi}~\tildeor~\interp{B}_{\phi}) = \interp{A \vee 
  B}_{\phi}$. Thus $\Gamma \in [A \vee B]$.  In a similar way,
$[B] \subseteq [A \vee B]$.
  
Conversely, let $\Gamma \in [A \vee B]$. Let $C$ such that $[A]
\cup [B] \subseteq [C]$, let $\phi$ be an assignment, $\sigma$ be a
substitution and $\Delta$ be $\Gamma$-adapted. By hypothesis,
$(\Delta \vdash \sigma (A \vee B)) \in
(\interp{A}_{\phi}~\tildeor~\interp{B}_{\phi})$. Let us consider the
three cases of Definition \ref{def:candidates} for $\tildeor$. First,
if $\Delta \vdash \sigma (A \vee B)$ has a neutral
cut-free proof. $\sigma A$ is $A$-adapted for $\phi, \sigma$ by the
first point of Proposition \ref{lem:adapt-composition}, so by
Proposition \ref{lem:adapt-composition} $(\Delta, \sigma A$) is $(\Gamma,
A)$-adapted. Since  $(\Gamma, A) \in [A] \subseteq [C]$ by Proposition
\ref{prop:basics2}, the sequent $\Delta, \sigma A \vdash \sigma C$ has
a cut-free proof by Proposition \ref{prop:basics2}. By similar arguments,
the sequent $\Delta, \sigma B \vdash \sigma C$ has a cut-free
proof. Hence, we can apply the $\vee$-elim rule on those three
premises and obtain a neutral cut-free proof of the sequent $\Delta
\vdash \sigma C$, which belongs to $\interp{C}_\phi$. Second, if
$(\Delta \vdash \sigma A) \in \interp{A}_{\phi}$. By Definition
\ref{def:adapted}, $\Delta$ is $A$-adapted, and since by Proposition
\ref{prop:basics2}, $A \in [A] \subseteq [C]$, we must have $\Delta
\vdash \sigma C \in \interp{C}_\phi$. The third and last case $(\Delta
\vdash \sigma B) \in \interp{B}_{\phi}$ is similar. In all three cases
we have $\Delta \vdash \sigma C \in \interp{C}_\phi$. Hence $\Gamma
\subseteq [C]$ for any $[C]$ upper bound of $[A]$, $[B]$ and it is an
element of their intersection {\em i.e.} of $[A]~\orc~[B]$.

\item Let us show $[A
  \Rightarrow B] \subseteq [A]~\impc~[B]$, which is by definition
  equivalent to $[A] \cap [A \Rightarrow B] \subseteq [B]$. Let
  $\Gamma \in [A] \cap [A \Rightarrow B]$, and $\Delta$ be
  $\Gamma$-adapted. Then $(\Delta \vdash \sigma A) \in \interp{A}_{\phi}$ and 
  $(\Delta \vdash \sigma A \Rightarrow \sigma B) \in \interp{A
    \Rightarrow B}_{\phi} = \interp{A}_{\phi} \tildeimp
  \interp{B}_{\phi}$.  If $\Delta \vdash \sigma A 
  \Rightarrow \sigma B$ has a neutral cut-free proof, since
  $\Delta \vdash \sigma A$ has a cut-free proof,
  $\Delta \vdash \sigma B$ has a neutral cut-free proof, and it
  belongs to $\llbracket B\rrbracket_{\phi}$.
  Otherwise $(\Delta \vdash \sigma A) \in \interp{A}_{\phi}$ and we
  apply Definition \ref{def:candidates} of $\tildeimp$ with an empty
  context $\Sigma$ to get $(\Delta \vdash \sigma B) \in
  \interp{B}_{\phi}$. Therefore, $\Gamma \in [B]$. \\

  Conversely let us show $[A] ~\impc~ [B] \subseteq [A \Rightarrow
    B]$. We have to prove that $[A \Rightarrow B]$ is an upper bound
  of the set of all the $c \in \Omega$ such that $c \cap [A] \subseteq
  [B]$. Let such a $c$, let $\Gamma \in c$, let $\phi$ be an assignment, 
  $\sigma$ be a substitution and $\Delta$ be a $\Gamma$-adapted
  context. We must show $(\Delta \vdash \sigma A \Rightarrow
  \sigma B) \in \interp{A \Rightarrow B}_{\phi} = \interp{A}_{\phi}
  \tildeimp \interp{B}_{\phi}$. 

  For this, let $\Sigma$ such that $(\Delta, \Sigma
  \vdash \sigma A) \in \interp{A}_{\phi}$. By Definition
  \ref{def:adapted}, $(\Delta, \Sigma)$ is $A$-adapted, and  by
  Proposition \ref{lem:adapt-composition} $(\Delta, \Sigma)$
  is $(\Gamma, A)$-adapted. From Proposition \ref{prop:basics2}, $[A]$
  and $c$ (by definition equal to some $\bigcap_{i \in \Lambda_c}
  [C_i]$) admit weakening, so $\Gamma, A \in c \cap [A] \subseteq
  [B]$. Therefore, by Definition \ref{mod:outer} of $[B]$,
  $\Delta, \Sigma \vdash \sigma B \in \interp{B}_\phi$, and
  the claim follows directly from Definition \ref{def:candidates} of
  $\tildeimp$.

       %

\item Let $\Gamma \in \bigcap \{ [(t/x)A], t \in \calT
  \}$. Let $\phi$ be an assignment and $\sigma$ be a substitution. Let
  $\Delta$ be $\Gamma$-adapted, we show that $\Delta \vdash 
  \sigma \forall{x} A \in \interp{\forall{x} A}_\phi$. We assume
  without loss of generality that $x$ does not appear in
  $\Delta$, nor in $\Gamma$, nor in $\sigma$. Let $t \in \calT$
  and $d \in M$. By freshness of $x$, $\Delta$ is also
  $\Gamma$-adapted for $\phi + (d/x)$, $\sigma + (t/x)$. Also, we have
  $(t/x) \sigma A =  (\sigma + (t/x)) A$ and by hypothesis, $\Gamma \in
  [(x/x) A] = [A]$. It means that $(\Delta \vdash (\sigma + (t/x)) A)
  \in  \interp{(x/x)A}_{\phi + (d/x)}$.
  Hence, by Definition \ref{def:candidates}, $(\Delta \vdash
  \forall{x} (\sigma A)) \in \tildefa \{ \interp{A}_{\phi +
    (d/x)} ~|~ d \in M \}$.

  Conversely, let $\Gamma \in [\forall{x} A]$. Let $t \in \calT$,
  $\phi$ be an assignment, $\sigma$ be a substitution and let
  $\Delta$ be $\Gamma$-adapted. Assume without loss of generality that
  $x$ does not appear in $\Delta$, nor in $\Gamma$, $\phi$,
  $\sigma$.

  By hypothesis $(\Delta \vdash \sigma \forall{x} A) \in
  \interp{\forall{x}A}_{\phi}$. If $\Delta \vdash \sigma \fa
  x A$ has a neutral cut-free proof then so does the sequent
  $\Delta \vdash (\sigma t / x)\sigma A)$. Since
  $\sigma (t/x)A = (\sigma t/x)\sigma A$, we have $(\Delta
  \vdash \sigma ((t/x)A)) \in \interp{A}_{\phi + (\interp{t}_\phi/x)}$, which
  is equal to $\interp{(t/x)A}_\phi$ by Proposition
  \ref{prop:substitution}.

  Otherwise, Definition \ref{def:candidates} ensures that for any $d$,
  including $\interp{t}_\phi$, $(\Delta \vdash
  \sigma ((t/x)A)) \in \interp{A}_{\phi  + (d/x)}$. Thus
  $(\Delta \vdash \sigma ((t/x)A)) \in \interp{(t/x)A}_{\phi}$. So
  $\Gamma \in [(t/x)A]$ for any $t$ and it is then an element of the
  intersection.


\item We first show that $[\exists{x} A]$ is an upper bound of the set
  $\{ [(t/x)A]~|~ t \in \calT \}$. Consider some  term $t$, and a
  context $\Gamma \in [(t/x)A]$. Let $\phi$ be an assignment, $\sigma$
  be a substitution and $\Delta$ a $\Gamma$-adapted context. Assume
  without loss of generality that $x$ does not appear in
  $\Delta$ nor in $\sigma$. Since $\sigma (t/x)A = (\sigma t/x)
  \sigma A$, we have by hypothesis $(\Delta \vdash (\sigma t/x)
  (\sigma A)) \in \interp{(t/x)A}_\phi$, which is equal to
  $\interp{A}_{\phi + (\interp{t}_\phi/x)}$ by Proposition 
  \ref{prop:substitution}. This shows that 
  $(\Delta \vdash \sigma \exists x A) \in \tilde{\exists} \{
  \interp{A}_{\phi + (d/x)}, d \in M \}$ by Definition
  \ref{def:candidates}. Hence $\Gamma \in 
         [\exists{x} A]$.  So $\exc~ \{ [(t/x)A]~|~t \in \calT \} 
         \subseteq [\exists{x} A]$.\\

  Conversely, let $\Gamma \in [\exists{x} A]$. Let $c =
  \bigcap[C_i]$ be an upper bound of $\{ [(t/x)A] ~|~ t \in
  \calT \}$.  We can choose $c = [C]$, since we
  need the intersection of the upper bounds. 

  Let $\phi$ be an assignment and $\sigma$
  be a substitution, let $\Delta$ be $\Gamma$-adapted and
  assume, without loss of generality, that $x$ does not appear in
  $C$, nor in $\Delta$ nor in $\sigma$. Finally, notice that $A
  \in [(x/x)A] \subseteq [C]$ and that by hypothesis on 
  $\Gamma$, $(\Delta \vdash \sigma \exists{x} A) \in
  \interp{\exists{x} A}_\phi$. 

  Assume $\Delta \vdash \sigma \exists{x} A$ has a neutral cut-free
  proof. Then, since $\sigma A$ is $A$-adapted, we have $(\sigma A
  \vdash \sigma C) \in \interp{C}_\phi$. In particular, by Proposition
  \ref{prop:basics}, this sequent has a cut-free proof. Since $x$ does
  not appear in $C$ nor in $\sigma$, we can apply an $\ex$-elimination
  rule between a proof of this sequent and the neutral cut-free proof
  of $\Delta \vdash \exists{x} \sigma A$, yielding a neutral
  cut-free proof of $\Delta \vdash \sigma C$. Hence $(\Delta \vdash
  \sigma C) \in \interp{C}_\phi$.
 
  Otherwise, by Definition \ref{def:candidates}, $\Delta
  \vdash \exists{x} \sigma A$ is such that for some term $t$ and element
  $d$, $(\Delta \vdash \sigma' A) \in \interp{A}_{\phi'}$, 
  calling $\sigma' = \sigma + (t/x)$ and $\phi' = \phi + (d/x)$. So,
  $\Delta$ is $A$-adapted for $\phi', \sigma'$. Since $A \in
  [C]$, this implies that $(\Delta \vdash \sigma' C) \in
  \interp{C}_{\phi'}$, but since $x$ does not appear in $C$, this is the
  same as  $(\Delta \vdash \sigma C) \in \interp{C}_\phi$.

  Therefore, $\Gamma \in [C]$. This is valid for any
  $[C]$ upper bound of $\{ [(t/x)A]~|~ t \in \calT \}$. So, $\Gamma$
  is in their intersection, that is $\tilde{\exists} \{ [(t/x)A], t
  \in \calT \}$.
\end{itemize}
}

\begin{proposition} \label{prop:gamma}
$\Gamma \in [\Gamma]$ 
\end{proposition}

\proof{Let $\Gamma = A_1, ..., A_n$. By Definition \ref{denotecontext}
  and Proposition \ref{prop:chech}, $[\Gamma] = [A_1 \wedge ... \wedge
A_n] = [A_1]~\andc~...~\andc~[A_n]$.  Using
Proposition \ref{prop:basics2}, we have $\Gamma \in
[A_1]~$, ..., $\Gamma \in [A_n]$, thus
$\Gamma \in ([A_1]~\andc~...~\andc~[A_n])$.}

\begin{definition}[The model $\calD$] \label{def:model-heyting}
Let $\calT$ be the set of classes of open terms modulo $\equiv$. Let
$\varphi$ be a substitution with values in $\calT$. For each function
symbol $f_i$ and each predicate symbol $P_j$ of the language we let:
\begin{itemize}
\item $\hat{f}_i: t_1, \cdots, t_n \mapsto f_i(t_1, \cdots, t_n)$
\item $\hat{P}_j: t_1, \cdots, t_n \mapsto [P_j(t_1, \cdots, t_n)]$
\end{itemize}
Let $\calD = \langle \calT, \Omega, \hat{f}_i, \hat{P}_j \rangle$
\end{definition}

\begin{proposition}[The model $\calD$]~ \label{prop:model-heyting}
\begin{itemize}
\item $\calD$ is an $\Omega$-structure in the sense of Definition
  \ref{def:Bstructure}.
\item the denotation $\interp{~}^\calD$ is such that for  any
  assignment ({\em i.e.} substitution) $\varphi$, any term $t$ and any
  proposition $A$:
  $$
  \interp{t}^\calD_\varphi = \varphi t \mbox{ and }
  \interp{A}^\calD_\varphi = [\varphi A]
  $$
\item $\calD$ is a model for $\equiv$ in the sense of Definition
  \ref{def:model}.
\end{itemize}
\end{proposition}
\proof{
The first point is immediate by Definition
\ref{def:Bstructure}. The proof of $\interp{t}^\calD_\varphi = \varphi
t$ is a straightforward structural induction on $t$. So is the proof
of $\interp{A}^\calD_\varphi = [\varphi A]$: the base case follows
from the definition of $\hat{P}_j$ and the inductive cases from
Proposition \ref{prop:chech}. Let us consider, for instance, the $\ex$
case. First assume that $x$ does not appear in $\varphi$, otherwise
rename it. Then, by definition $\interp{\ex{x} A}^\calD_\varphi = \exc \{
\interp{A}^{\calD}_{\varphi+(t/x)}, t \in \calT \}$. By induction
hypothesis, this is equal to $\exc \{ [(\varphi + (t/x))A], t \in \calT
\} = \exc \{ [(t/x)(\varphi A)], t \in \calT \}$. By Proposition
\ref{prop:chech}, this is equal to $[\ex{x} (\varphi A)] =
    [\varphi(\ex{x} A)]$.

The last point is a direct consequence of Proposition
\ref{prop:basics2} and of the second point: if $A \equiv B$ then, for
any substitution $\varphi$, $\varphi A \equiv \varphi B$ and
$\interp{A}^\calD_\varphi = [\varphi A] = [\varphi B] =
\interp{B}^\calD_\varphi$.
}

\subsection{Cut Elimination}

First, with the help of the model $\calD$ we conclude directly that
the cut-free calculus is complete:

\begin{proposition}[Completeness of the cut-free calculus]
\label{prop:strongcompleteness}
If the sequent $\Gamma \vdash B$ is valid  in the model $\calD$ ({\em
  i.e.} $\topc \subseteq \interp{\Gamma \vdash B}$ or equivalently
$\interp{\Gamma} \subseteq \interp{B}$), then it has a cut-free proof.
\end{proposition}

\proof{Let $\varphi$ be the identity assignment. By Proposition
  \ref{prop:gamma} and by hypothesis: 
$$
\Gamma \in [\Gamma] = \interp{\Gamma}_\varphi \subseteq \interp{B}_\varphi = [B]
$$
By Proposition \ref{prop:basics2}, the sequent $\Gamma \vdash B$ has a
cut-free proof.}

\begin{theorem}[Cut elimination]\label{prop:cutelim-norm}
If the sequent $\Gamma \vdash B$ is provable, then it has a cut-free
proof.
\end{theorem}
\proof{From the soundness theorem (Proposition \ref{soundcomplete})
  and Proposition \ref{Heyting},  if $\Gamma \vdash B$ is provable,
then it is valid in all Heyting algebra-valued models of the
  congruence, in particular $\calD$. Hence, by
  Proposition \ref{prop:strongcompleteness}, it has a cut-free proof.}

\begin{remark}
In the previous proof, the induction is performed by the
soundness theorem, while the inductive cases are performed
by Proposition \ref{prop:chech}, that ensures that $[\_]$ is a
model interpretation. So, we observe a split of the cut-elimination
theorem in two parts. This has to be compared to proofs of cut
elimination {\em via} normalization, that, given a proof of $\Gamma
\vdash A$, would show directly $[\Gamma] \subseteq [A]$ or something
similar (Theorem 3.1 of \cite{GDowBWer03} for instance). This split is
essentially made possible by Definition \ref{mod:outer}.
\end{remark}

\section{Application To Simple Type Theory} \label{sec:disc}
As a particular case, we get a cut elimination proof for simple
type theory. 

Let us inspect the model construction in more details in this
case. Based on the 
language of simple type theory, we first build the truth values
algebra of sequents $\calS$ of Definition
\ref{def:candidates}. Then using the super-consistency of simple type
theory, we build the model $\calM$ as in Proposition \ref{stt}. In
particular $M_{\iota} = \{0\}$, $M_{o} = S$ (see Definition
\ref{def:candidates}), and $M_{T \rightarrow U} = M_U^{M_T}$.  Then,
we build the model $\calD$ as in Proposition \ref{prop:model-heyting}
and we let $D_T = \calT_T$, where $\calT_T$ is the set of equivalence
classes modulo $\equiv$ of terms of sort $T$.  In particular, we have
$D_{\iota} = \calT_\iota$ and $D_o = \calT_o$.

This construction differs from that of the $V$-complexes of
\cite{PRA,TAK,AND,LIP03,OHerJLip10,OHerJLip08b} used to prove cut
admissibility in higher-order logic. Let us analyze this further.

\subsection{Principles Of The Proof with $V$-Complexes}

We here give a sketch of a proof with $V$-complexes in the simpler
case of classical logic, as given in \cite{AND} for instance or, in a
modern and intuitionistic version, in \cite{LIP03}. Notice that, in
contrast with the presentation of Definition \ref{def:stt}, the
$\varepsilon$ symbol is absent so that the logical connectors merge
with the associated ``dotted'' constant. For instance, in this section
we shall consider that $\dot{\wedge}$ is the same as $\wedge$.\\

Let $\Gamma \vdash \Delta$ be a sequent that has no proof in the
cut-free sequent calculus. We assume given a semi-valuation $V$
\cite{Schutte} compatible with this sequent {\em 
  i.e.} a partial interpretation function from the propositions into
$\{ 0,1 \}$ such that $V(\Gamma) = 1$ and $V(\Delta) = 0$. Such a
semi-valuation can be obtained by an (infinite) tableau procedure or
as an abstract consistency property, see \cite{AND,LIP03}. It is
weaker than a model interpretation, in the 
sense that it is partial, and consequently consistency conditions are
weaker: if we know the truth value of a proposition $A$,
enough must be known on the truth value of its immediate
sub-propositions. For instance, $V(A \wedge B) = 0$ {\em implies}
$V(A) = 0$ or $V(B) = 0$ and the other value might be left undefined.

The goal, and the difficulty in the Simple Type Theory case - or
higher-order logic - as identified by Sch\"{u}tte
\cite{Schutte}, resides in the extension of $V$ into a model
interpretation. The answer, given independently by Takahashi
\cite{TAK} and Prawitz \cite{PRA}, is to construct a new
interpretation domain, called $V$-complexes, as follows.\\

First,  for every type $T$, the interpretation domain of the model is
built by glueing together a syntactic and a semantic part:
$$
V_T = \calT_T \times \calM_T
$$
where $\calT_T$ is the set of terms of type $T$ that are in normal
form, $\calM_\iota = \{ \iota \}$, $\calM_{o} = \{ 0, 1 \}$  and
$\calM_{A \rightarrow B}$ is the function space $V_B^{V_A}$,
{\em i.e.} it is composed of functions of $\calT_A \times \calM_A
\rightarrow \calT_B \times \calM_B$ that verify the following
criterion. A pair $\langle t, f \rangle$ belongs to $V_T$ if and only
if $t$ is in normal form and of type $T$, and:
\begin{itemize}
\item when $T$ is $\iota$, $f$ is equal to $\iota$
\item when $T$ is $o$ and $V(t)$ is defined, $f$ is
  equal to $V(t)$ (this way we enforce the adequacy with the
  semi-valuation $V$). Otherwise $f$ can be either $0$ or $1$, and
  indeed both $V$-complexes $\langle t, 0 \rangle$ and $\langle
  t, 1 \rangle$ belong to the domain $V_o$.
\item when $T$ is a function type $A \rightarrow B$, $f \in \calM_{A
  \rightarrow B}$ can be decomposed in a function $f_1$  from $\calT_A
  \times \calM_A$ to $\calT_B$ and a function $f_2$ from $\calT_A
  \times \calM_A$ to $\calM_B$. Given any $V$-complex $\langle
  t',a\rangle \in \calM_A$, we require that $f_1(\langle t',a\rangle) =
  nf(t~t')$ and $\langle nf(t~t'), f_2(\langle t', a \rangle) \rangle
  \in V_B$, where $nf(t~t')$ is the normal form of $t~t'$.
\end{itemize}

$V$-complexes were introduced to deal with two main problems of
higher-order logic: impredicativity and intensionality. Tait's method
\cite{WTai66} solves 
the first problem by performing an induction on the type, this way
avoiding an impossible induction on term size. This has to be
improved to handle intensionality: logically speaking $\top$ and
$\top \and \top$ must have the same denotation, while we must still be
able to do a semantic distinction between the denotations of $P(\top)$
and $P(\top \land \top)$ since the first propositions are equiprovable
while the second are not. Moreover, the
interpretation $\langle P, f \rangle$ of $P$ must be such that the
logical denotation (the second component of the interpretation) of
$P(\top)$ and $P(\top \wedge \top)$ are different. This is possible
only if $f_2$ uses both sides of its argument, in particular we must
have:

$$
f_2(\langle {\top} {\wedge} {\top}, \llbracket {\top}
{\wedge} {\top} \rrbracket \rangle) \neq f_2(\langle
{\top}, \llbracket {\top} \rrbracket \rangle)
$$
although $\llbracket {\top} {\wedge} {\top} \rrbracket =
\llbracket {\top} \rrbracket = 1$.

This is achieved by introducing a syntactical component into the
semantic denotation of the terms. It then becomes possible to have
different values and behaviors, depending on this syntactical
component. This is reflected by the behavior of the function $f_2$: it
crucially depends on both components.

As a consequence, we separate the logical denotation of terms, (that
equalizes $\top$ and $\top \land \top$ in any Boolean algebra or
Heyting algebra), from their interpretation in the model (that does
not and that is more related to the meaning of the proposition that to
its denotation) lying at a lower level, the level of $V$-complexes.\\

On the basis of $V$-complexes, we define an interpretation for any
term $t$ by induction on its structure. Let us see some key cases:
\begin{itemize}
\item if $t$ is not a logical symbol, we interpret it
  by a default $V$-complex associated to $t$, of shape $\langle t,
  d\rangle$ (of course, a lemma states that it exists).
\item if $t$ is the logical symbol $\dot{\fa}_T$ we construct the
  $V$-complex $\langle \dot{\fa}_T, f \rangle$ where $f$ is the
  following function: to any $V$-complex $\langle t, g \rangle$ of
  type $T \rightarrow o$ it associates the $V$-complex of type $o$
  $\langle \dot{\fa}_T~t, v \rangle$ where $v$ is equal to $1$ if and
  only if for any $V$-complex $d$ of type $T$, $g(d)$ is a $V$-complex
  that has $1$ as second component. So that we quantify over {\em all}
  the $V$-complexes of type $T$.
\item the interpretation of the application symbol $\alpha$ applies a
  $V$-complex $\langle t, f \rangle$ to another one $\langle u, g
  \rangle$ as $f(\langle u, g \rangle)$. Notice that its first member
  is, by the conditions on $f$, the normal form of $(t~u)$.
\end{itemize}

It is a matter of technique to check that this construction
really produces $V$-complexes. The last step is to consider a
generalized notion of model, since now terms of type $o$ have a
denotation in $\calT \times \{ 0, 1 \}$ which is not a Boolean
algebra. Then we can state that the interpretation we built is
compatible with $V$, and the propositions of $\Gamma$ are interpreted
by $1$ (as a second component) while those of $\Delta$ are interpreted
by $0$. Therefore the sequent $\Gamma \vdash \Delta$ is not
valid if it has no cut-free proof. This yields a proof of a strong
version of the completeness theorem from which we derive the
cut-elimination theorem.

\subsection{Comparison}

In contrast, in our construction, we have two separate models, the
term model $\calD$, that corresponds to the left-hand side of a $V$-complex, and
the model $\calM$ that corresponds to the right-hand part.

The novelty is that  $\calM_{A \rightarrow B}$ is just $\calM_A
\rightarrow \calM_B$ and not $\calT_A \times \calM_A \rightarrow
\calM_B$. This is possible because when we build $\calM$, instead of
taking  $\calM_o = \{0,1\}$, we have taken $\calM_o = S$ that is a
truth value algebra but not a Heyting algebra. Thus $\llbracket\dtop
\dand \dtop \rrbracket^\calM$ and $\llbracket \dtop \rrbracket^\calM$
need not be equal, the truth values containing more information,  
and we do not need to glue an extra syntactical argument $\dtop \dand
\dtop$ or $\dtop$ to have $f_2(\interp{\dtop \dand \dtop}) \neq
f_2(\interp{\dtop})$.

The same phenomenon arises in $\calD$ since we choose a syntactic
model: following Proposition \ref{prop:model-heyting} $\llbracket\dtop
\dand \dtop \rrbracket^\calM$ and $\llbracket \dtop \rrbracket^\calM$
are respectively equal to $\dtop \dand \dtop$ and $\dtop$. In a
similar way $\interp{P(\dtop)}^\calD = \interp{P(\dtop)}^\calM 
\lhd P(\dtop) = [P(\dtop)]$, and from Proposition \ref{prop:basics2}
$\interp{P(\dtop)}^\calD$ contains $P(\dtop)$ , while $\interp{P(\dtop
  \dand \dtop)}^\calD$ contains $P(\dtop \dand \dtop)$. None of those
interpretations, in the general case, contains the other proposition,
therefore they are not equal, as required.\\

One may also wonder where the separation of the logical denotation $1$
of $\dtop$ from its interpretation in $\calD$ appears in our
proof. The expression $\dtop$ has an existence only at the {\em term}
level since at the {\em propositional} one it is replaced by
$\varepsilon(\dtop)$. The interpretations in $\calD$ of $\dtop$ and
$\varepsilon(\dtop)$ respectively correspond to the interpretation in
$V_o$ and the denotation in a Heyting algebra of $\top$ in previous
proofs. 

The separation between denotation and interpretation, that had to be
defined ``by hand'', introducing a new definition for models, in the
earlier works with $V$-complexes
\cite{PRA,TAK,AND,LIP03,OHerJLip10,OHerJLip08b}, is automatically 
captured by the simple syntactical device $\varepsilon$.

\subsection{Conclusion}

Thus the main difference between our model construction and that
of the V-complexes is that we have broken this dependency on $u$ of
the right component of the pair obtained by applying $\langle t, f
\rangle$ to $\langle u, g \rangle$. This leads to a
two-stage construction where the very notion of $V$-complex has
vanished and the second model is syntactical. The reason why we 
have been able to do so is that starting with an underlying model of
sequents $S$, our semantic objects $[A]$ are much sharper, and do
not require additional construction. Moreover, the presence of the
symbol $\varepsilon$ has simplified the dependency of the semantics on
the syntax and allowed a purely syntactical model at the term-level.\\

It has to be noticed that super-consistency allows us to construct a
model on more usual Heyting algebras, such as the Lindenbaum Heyting
Algebra, or the context-based ones used for cut elimination
\cite{OHerJLip08b,MOka02}, where $[A]$ is defined as the set of
contexts $\Gamma$ such that $\Gamma \vdash A$ has a cut-free proof. It
gives us an interpretation on this algebra that satisfies the
congruence $\equiv$. The pitfall is that if we build such a model in
ordinary way, we cannot prove that $\interp{A} = [A]$. To achieve this
goal we have to proceed by first defining
the algebra of sequents in an untyped way, and then by extracting the
needed contexts in order to force $\interp{A}$ to be equal to
$[A]$. So a two-stage construction seems unavoidable when one uses
super-consistency to show the admissibility of the cut rule.

It remains to be understood if such a construction can also be
carried out for a normalization proof.

\bibliographystyle{plain}
\bibliography{cuts}

\end{document}